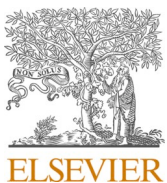
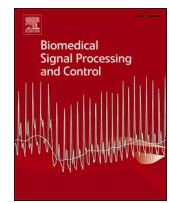

# Sentiment analysis in non-fixed length audios using a Fully Convolutional Neural Network

María Teresa García-Ordás [a], Héctor Alaiz-Moretón [a], José Alberto Benítez-Andrades [b],[*], Isaías García-Rodríguez [a], Oscar García-Olalla [c], Carmen Benavides [b]

[a] *SECOMUCI Research Group, Escuela de Ingenierías Industrial e Informática, Universidad de León, Campus de Vegazana s/n, C.P. 24071 León, Spain*
[b] *SALBIS Research Group, Department of Electric, Systems and Automatics Engineering, Universidad de León, Campus of Vegazana s/n, León, 24071 León, Spain*
[c] *Artificial Intelligence Department, Xeridia S.L., Av. Padre Isla 16, 24002 León, Spain*

ABSTRACT

In this work, a sentiment analysis method that is capable of accepting audio of any length, without being fixed a priori, is proposed. Mel spectrogram and Mel Frequency Cepstral Coefficients are used as audio description methods and a Fully Convolutional Neural Network architecture is proposed as a classifier. The results have been validated using three well known datasets: EMODB, RAVDESS and TESS. The results obtained were promising, outperforming the state-of-the-art methods. Also, thanks to the fact that the proposed method admits audios of any size, it allows a sentiment analysis to be made in near real time, which is very interesting for a wide range of fields such as call centers, medical consultations or financial brokers.

## 1. Introduction and related work

Speech emotion recognition is a significant problem as well as a challenge, because of its numerous applications, such as audio surveillance, E-learning, clinical studies, lie detection, entertainment, computer games and call centers [1]. Emotion processing is also important for polarity detection which is very useful in social events, political movements and marketing campaigns [2]. Emotions play an important role in our life, not only in human interaction, but also in decision-making processes, and in the perception of the world around us [3]. Language, speech, is the way humans communicate. Humans also communicate regularly by writing, and in this case, sentences can be misinterpreted if the tone of the speaker is not known. Since emotions help us to understand each other better, it is necessary to extend this understanding to computers.

Most of the work that is normally carried out into sentiment classification analyzes texts using natural language processing techniques. Numerous examples have appeared over the years. For example, in [4], the authors analyze the mood of society on a particular news item from Twitter posts using NLP (Natural Language Processing) techniques such as semantics and Word Sense Disambiguation. A similar approach is developed in [5]. In this case, the objective is to use a data mining approach on text-feature extraction, classification, and dimensionality reduction, using sentiment analysis to analyze and visualize the opinion of twitter users. After the dataset collection stage, the preprocessing stage, and the NLP step, the classification process is carried out using Bayesian Neural Networks. In the case of [6], the main purpose is to develop an application capable of analyzing the content of online courses and the contributions of their learners such as video transcriptions, readings, questions and answers of the evaluation activities, posts in forums, etc. using NLP, to improve the teaching material and the teaching–learning processes of these courses.

Recent techniques based on deep neural networks have demonstrated a good performance in text emotion recognition. In [7], Bidirectional Encoder Representations from Transformers (BERT) are used for sentiment classification outperforming other methods in the literature and solving the problems of traditional sentiment analysis methods, such as the need for complex feature engineering. More recently, in [8], a stacked ensemble method for predicting the degree of intensity for emotions were proposed. In [9], another ensemble of symbolic and subsymbolic tools named SenticNet 6 was presented. In the same year, [10] proposed a Bidirectional Emotional Recurrent Unit (BiERU) for conversational sentiment analysis which takes context information into account in dialogues. In 2021, [11], an Attention-based Bidirectional

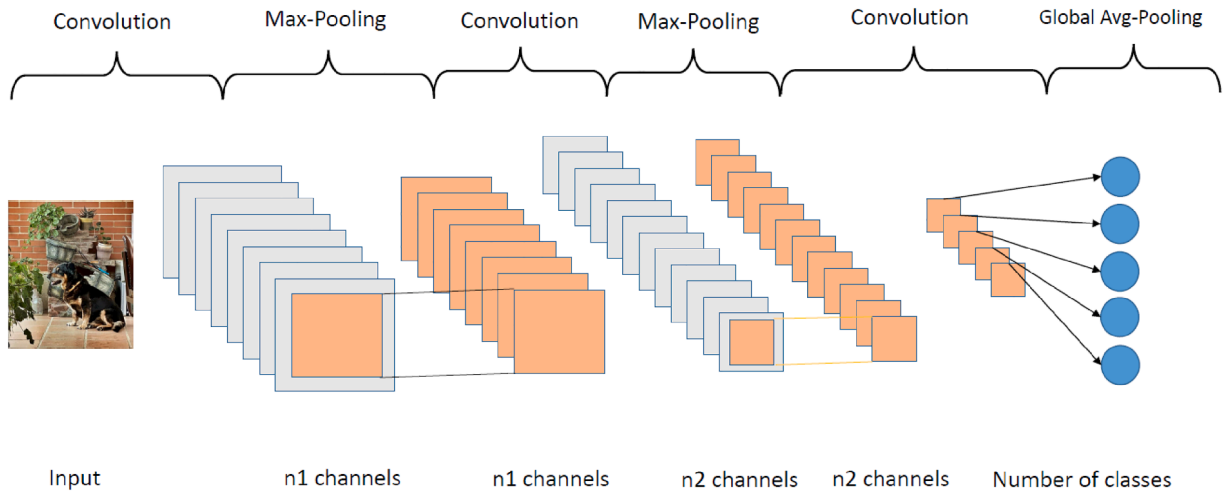

Fig. 1. Vanilla FCN representation.

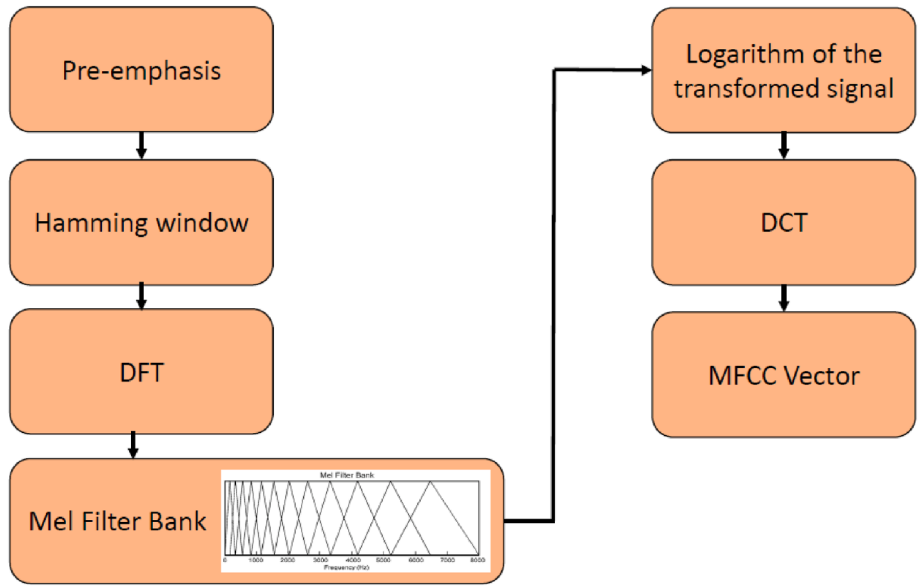

Fig. 2. MFCC extraction process.

CNN-RNN Deep Model (ABCDM) was developed. This method utilizes two independent bidirectional LSTM (Long Short Term Memory) and GRU (Gated recurrent unit) layers to extract both past and future contexts.

If it is wished to analyze the audio, as is the case of this study, the audio file must be transcribed to use all of the studied NLP techniques for emotion recognition throughout the extracted text.

In recent years, techniques that use the audio signal have been developed with less necessity for NLP. This has the advantage that the recognition is invariant to the language and is also able to detect sarcasm or irony. For dealing with audio files, there are many techniques for extracting features.

Mel-frequency cepstral coefficients (MFCCs) are coefficients derived from a type of cepstral representation of the audio clip (a nonlinear "spectrum-of-a-spectrum") and it is the most used representation of an audio signal [12–14]. According to [12] and using their hierarchical sparse coding (HSC) scheme, pitch, intensity, low-pass intensity, high-pass intensity and the norm of the absolute vector derivative of the first 10 MFCC components are the most representative features for audio description.

By combining MFCC, discriminant analysis and NSL (Neural Structure Learning) [13] superior recognition rates are generated compared to other traditional approaches such as MFCC-DBN (Deep Belief Network), MFCC–CNN (Convolutional Neural Network), and MFCC-RNN (Recurrent Neural Network) during the experiments on an emotion dataset of audio speeches. The method proposed by Uddin et al. [13], can be adopted in smart environments such as homes or clinics to provide affective healthcare. Since NSL is fast and easy to implement, it can be tried on edge devices with limited datasets collected from edge sensors.

MFCC features are also used in [14]. A combination of spectral features was extracted from the audio files, which are further processed and reduced to the required feature set. The classification step was carried out using a bagged ensemble made up of support vector machines with a Gaussian kernel.

It frequently happens that so many features are extracted from the audio files that a lot of them are not relevant and only produce errors in the classification process. Therefore finding an appropriate feature representation for audio data is central to speech emotion recognition. The number of features obtained by acoustic analysis reaches very high values depending on the number of acoustic parameters used and statistical variations of these parameters. For all of these reasons, many

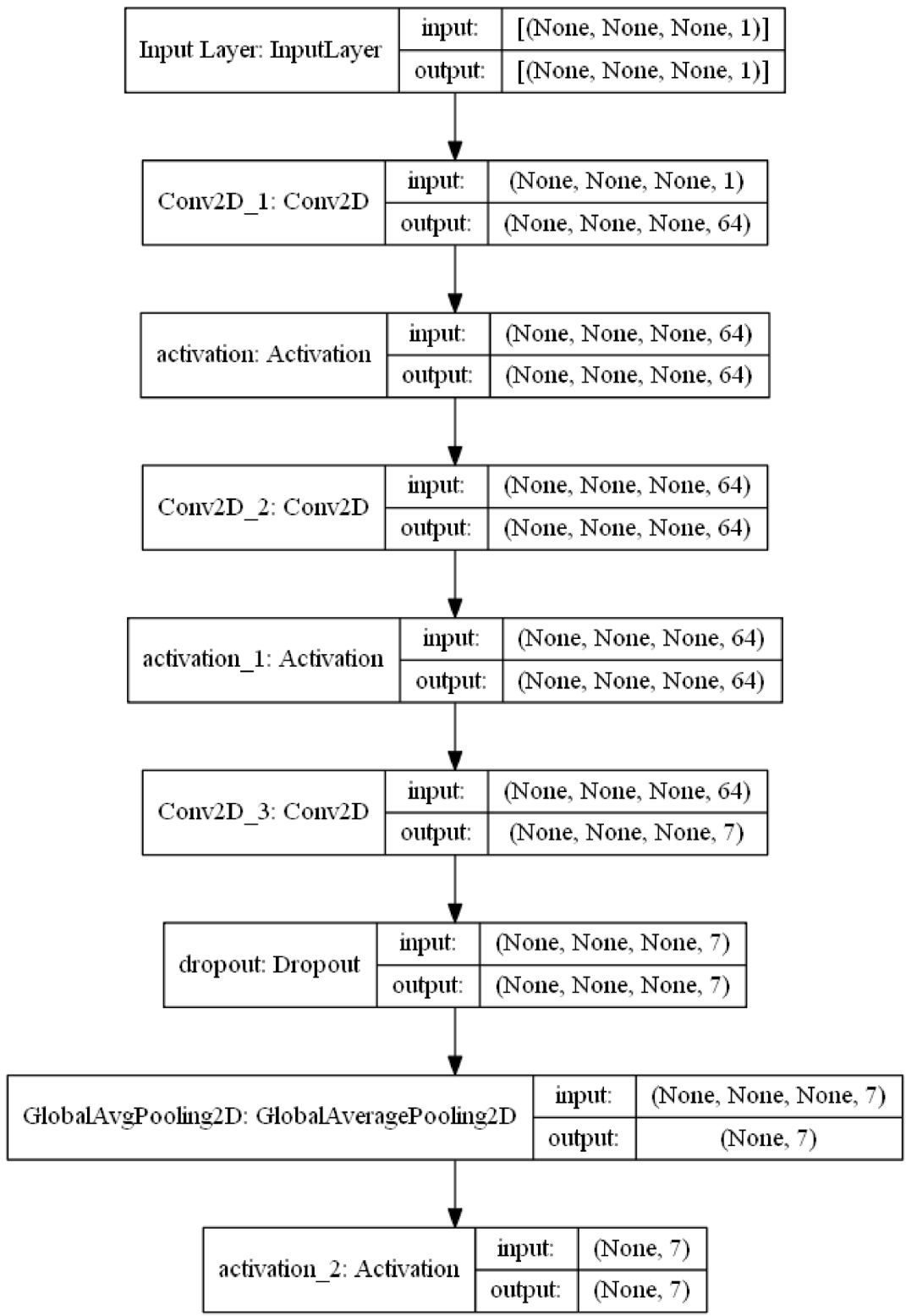

**Fig. 3.** FCN representation.

feature reduction and feature selection techniques began to emerge.

In [15], LDA (Linear Discriminant Analysis) and PCA (Principal Component Analysis) techniques are used to reduce the number of extracted features and after this process, a genetic algorithm carries out the classification step. The results obtained in this work are promising obtaining an accuracy of 90.28% for the CASIA dataset, 76.40% for the SAVEE dataset, and 71.05% for the FAU Aibo dataset for speaker dependent and 38.55% (CASIA), 44.18% (SAVEE), and 64.60% (FAU Aibo) for speaker-independent (SI).

Jing et al. [16] also try to find out the best characteristics for rendering an audio file. A novel type of feature related to prominence is proposed, which, together with traditional acoustic features, are used to classify seven different typical emotional states. The prominence features are validated through speaker-dependent and speaker-independent experiments with Support Vector Machines (SVM), k nearest neighbors (kNN), Naive Bayes (NB) and Partial Least Squares-Discriminatory Analysis classifiers. The results show that the average recognition rate achieved using the combined features is improved by 6% in speaker-dependent experiments and by 6.2% in speaker-independent experiments compared with that achieved using only acoustic features.

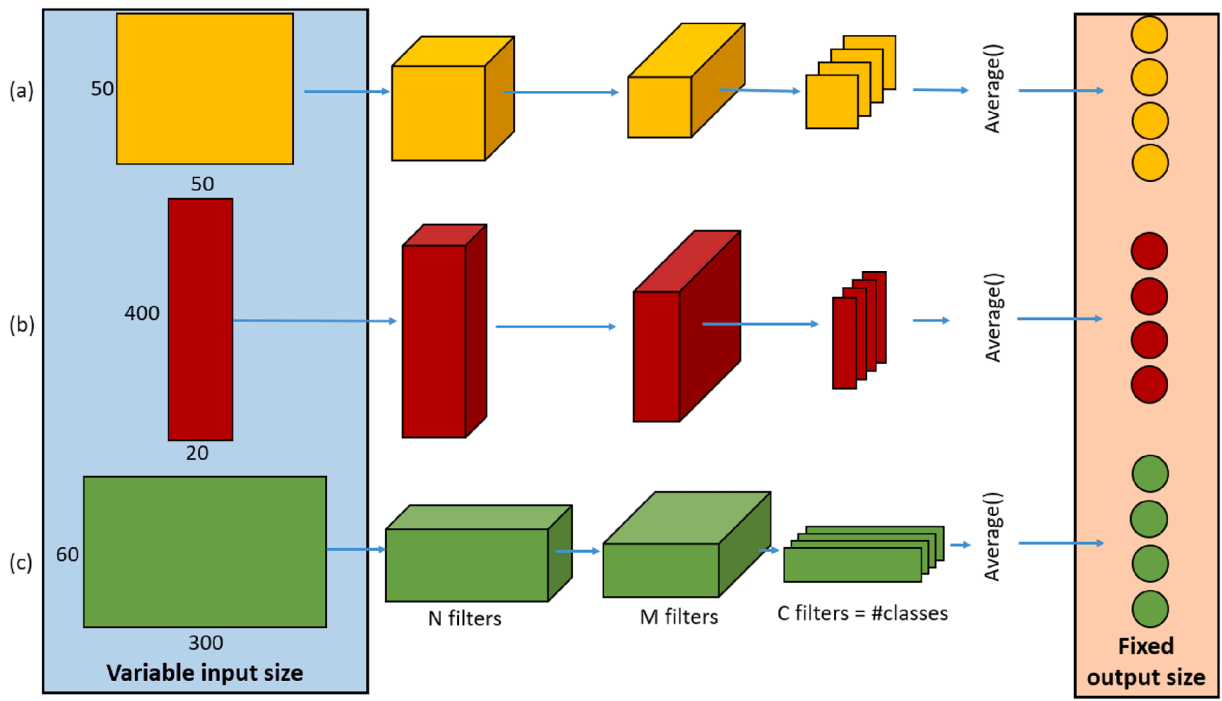

**Fig. 4.** FCN example for different input shapes. The output size remains constant regardless the original size of the input data.

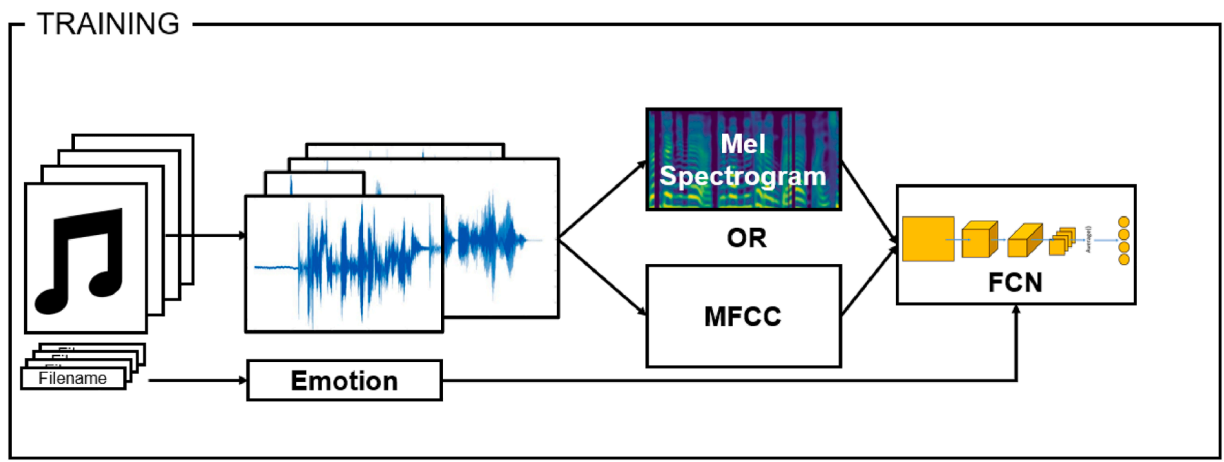

**Fig. 5.** Flowchart of the training process. In this case, the emotion of each file is noted in the filename.

Mutual information is also used to determine the minimum number of features required to recognize and analyze the audio [17]. In this case, after feature extraction, the authors used Gaussian Mixture Models (GMM) as a classifier on the EMODB database.

In [18], a new statistical feature selection method is proposed based on the changes in emotions on acoustic features since not all of these features are effective for emotion recognition. Furthermore, different emotions may affect different vocal features. In this case, kNN, is used as classifier. The feature size of 1,582 was reduced to between 77% and 86% with PCA, between 95.6% and 98.6% with SFS (Sequential Forward Selection), between 96.5% and 98.8% with FCBF (Fast Correlation-Based Filter) and between 75% and 98.6% with the method proposed by Ozseven et al., depending on the dataset.

These audio features are usually used to feed different machine learning or deep learning techniques for speech emotion recognition tasks.

For example a multilingual speech emotion recognition is proposed in [19]. It was tested on Japanese, German, Chinese, and English emotional speech corpora. The recognition performance was enhanced and examined using cross-speaker and cross-corpus evaluation, obtaining promising results even with a different speaker or language.

In [20], a data augmentation method called Segment Repetition based on High Amplitude (SRHA) was proposed to solve the lack of data problem using Long Short-Term Memory (LSTM) for classification.

In [21], a deep neural network with kernel extreme learning was carried out to classify EmoDB and IEMOCAP datasets.

In [22], an autoencoder architecture is proposed to encode audio in order to improve the performance of speech emotion recognition systems.

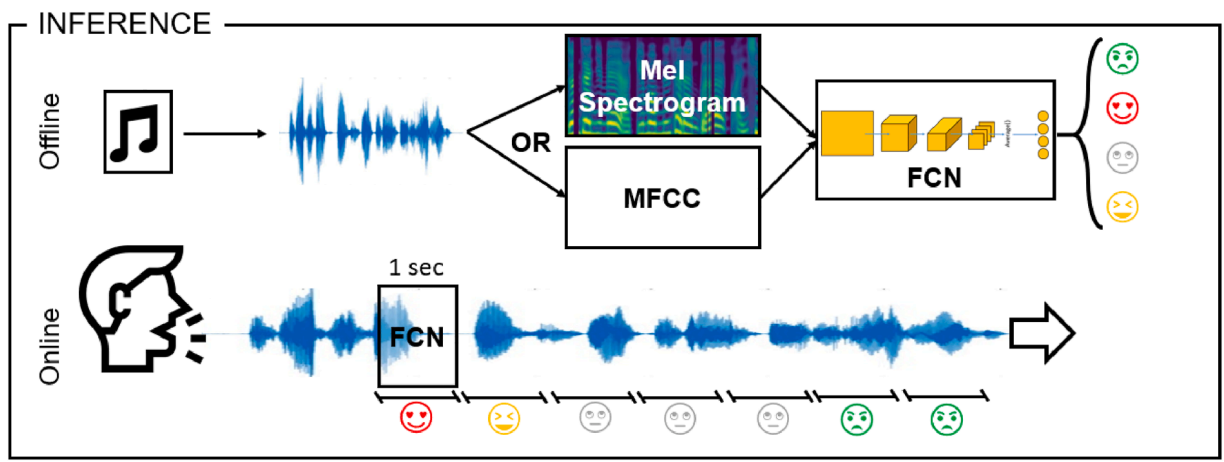

**Fig. 6.** Flowchart of the inference step. First, the offline process when an audio file is used as input data. Second, the online process when real-time audio is processed to determine emotion every second.

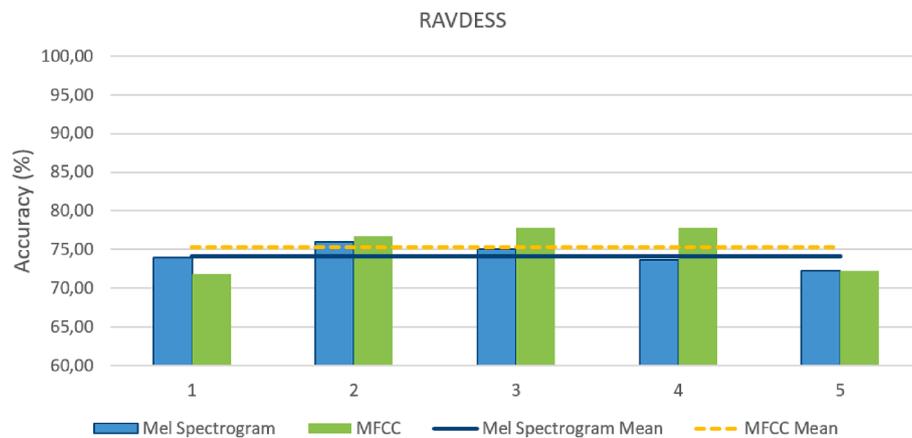

**Fig. 7.** RAVDESS accuracy for each of 5 test subsets and their mean using Mel Spectrogram and MFCC audio descriptions.

**Table 1**
RAVDESS Results summary.

| Audio Desciptor | Mean | Median | Std |
| --- | --- | --- | --- |
| Mel Spectrogram | 74.17 | 73.96 | 1.4435 |
| MFCC | 75.28 | 76.74 | 2.9871 |

Most of the recent research uses convolutional networks with dense layers to carry out a sentiment analysis [23–28]. For this reason, audio files have to be of a fixed size.

In our work, a method capable of analyzing audio files of any size is proposed, without the need for that size to be fixed previously, using an architecture based on Fully Convolutional Neural Networks (FCN). FCN have been used in different areas for the resolution of many problems. In [29], a Unet architecture and a fully convolutional layer as a bottleneck between encoder-decoder blocks are used to carry out an automated nuclei segmentation for diagnosing cervical pre cancerous and cancerous lesions. In [30–32], FCN are used for the segmentation of the retina or the retinal fluid. A multi-scale recurrent fully convolution neural network to identify and segment laryngeal leukoplakia lesions is proposed in [33]. FCN are also applied in tooth segmentation problems [34], brain tumor segmentation [35,36], ventricle segmentation [37], myocardial segmentation [38], amongst others.

In addition to being used for segmentation, in [39], FCN are used to detect new T2-w lesions in longitudinal brain MR images which play an important role in multiple sclerosis (MS) diagnosis and follow-up, or to detect abnormalities, such as polyps, ulcers and blood, in gastrointestinal (GI) endoscopy images [40].

In the field of optical character recognition (OCR), Ptcha et al. [41], present a fully convolutional network architecture which outputs arbitrary length symbol streams from handwritten text.

In [42], a FCN is used as an automatic feature extraction method in audios for emotion recognition. More recently, in [43], a new model architecture which is composed of an attention-based convolution long short-term memory neural network and a fully connected neural network was proposed for emotion recognition.

In our work, an FCN architecture is proposed for classifying audio files of any length and being able to carry out a near real time emotion identification.

The rest of the paper is organized as follows: In Section 2, the datasets employed and the proposed methodology are described. The different experiments carried out, are summarized and discussed in Section 3 and finally, the conclusions are detailed in Section 4.

## 2. Methods

### 2.1. Datasets

The proposal has been evaluated on three well-known datasets: RAVDESS [44], EMODB [45] and TESS [46]. These three datasets are widely employed by researchers in emotion recognition.

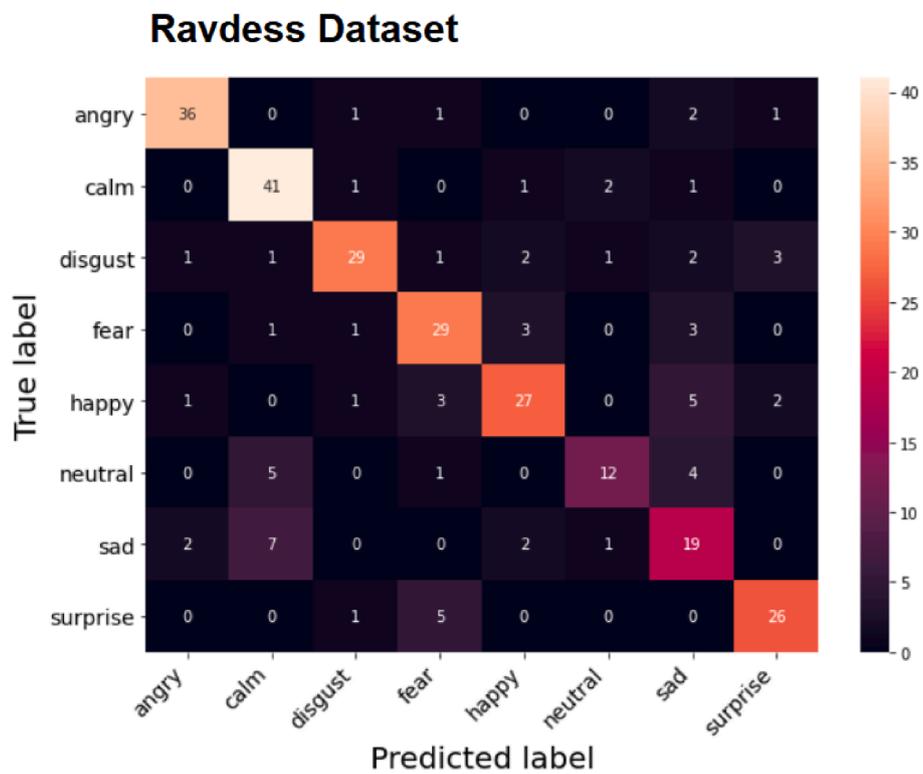

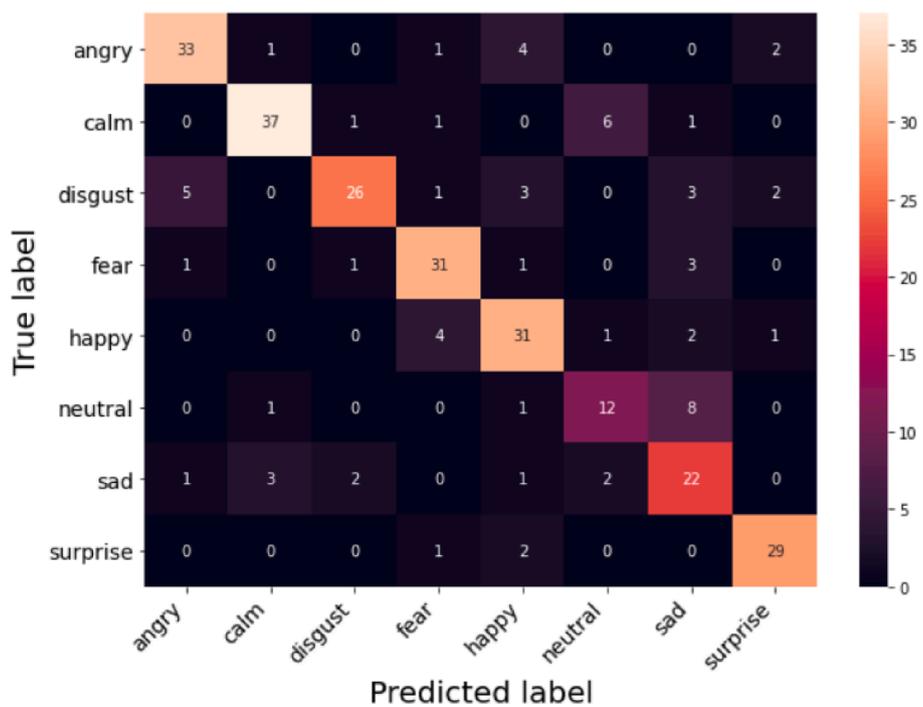

**Fig. 8.** RAVDESS confusion matrix for the best training with (a) mel spectrogram audio description and (b) MFCC audio description.

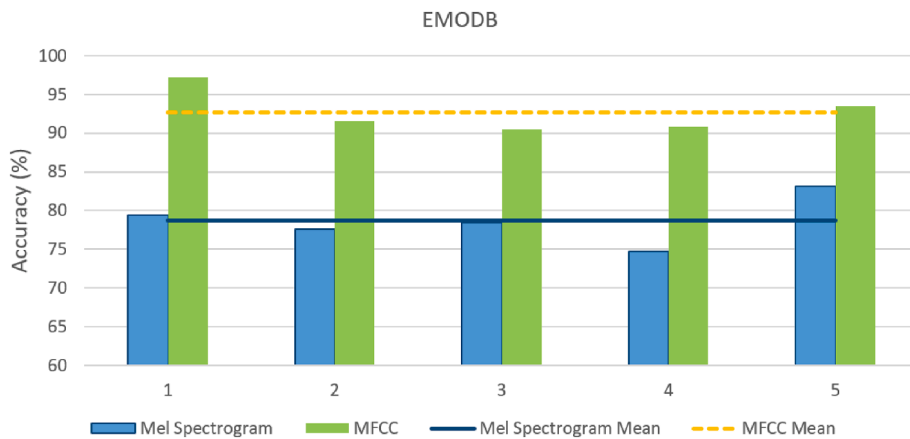

**Fig. 9.** EMODB accuracy for each of 5 test subsets and their mean using Mel Spectrogram and MFCC audio descriptions.

**Table 2**
EMODB Results summary.

| Audio Descriptor | Mean | Median | Std |
| --- | --- | --- | --- |
| Mel Spectrogram | 78.69 | 78.50 | 3.0601 |
| MFCC | 92.714 | 91.59 | 2.7515 |

*2.1.1. RAVDESS*

The Ryerson Audio-Visual Database of Emotional Speech and Song (RAVDESS) dataset [44], is made up of audio files and the visual English recordings of 24 people (12 male and 12 female). RAVDESS contains 7,356 files (total size: 24.8 GB). In this case, only the speech samples are used and the sentences are pronounced with eight different emotional expressions, thus the classes for our problem are as follows: sad, happy, angry, calm, fearful, surprised, neutral and disgust.

*2.1.2. EMODB*

The EMODB dataset [45] contains 535 German audio files divided into seven emotion classes: anger, sadness, fear or anxiety, neutral, happiness, disgust, and boredom. Ten actors (5 female and 5 male) simulated the emotions, producing 10 German utterances (5 short and 5 longer sentences) which could be used in everyday communication and are interpretable in all applied emotions. The dataset was evaluated in a perception test on the recognizability of emotions and their naturalness. Utterances were judged as natural by more than 60 The data was phonetically tagged with special markers for voice quality, phonatory and articulatory settings, and articulatory characteristics.

*2.1.3. TESS*

The Toronto Emotional Speech Set (TESS) [46] is an acted dataset comprising 2,800 short audios spoken by two actresses of 20 and 60 years old respectively. Each of them simulated seven sentiments for 200 different sentences. The dataset was blindly labeled by a group of 56 students. The emotions of the dataset are as follows: happy, sad, fear, surprised, angry, disgusted and neutral.

*2.2. Fully Convolutional Neural Network*

Traditionally, convolutional neural networks are made up of several convolutional layers and some fully connected layers at the end to carry out the classification or regression. This architecture forces us to determine the size of the input.

In this paper, a Fully Convolutional Neural Network (FCN) is proposed as a classifier. An FCN is a Convolutional Neural Network but without fully connected layers.

Fully Convolutional Networks (FCN), only use convolutional layers and upsampling or downsampling ones. Convolutional layers learn a fixed number of values determined by the size of the filters instead of the size of the input data. Thanks to this, FCN enable us to use variable input data which is essential for the correct treatment of conversational audio files.

Furthermore, FCN enable us to evaluate our full sentiment audio file or split it in smaller audio files to carry out an almost "real time" tracking of the sentiments or even distinguish the sentiments of every speaker in a conversation independently of the length of each utterance.

FCN is made up of convolutional layers. They usually also include regularization layers such as Dropout or BatchNormalization. The key point of FCN to enable variable input size lies in the use of a global pooling layer at the end. There are two main global pooling layers: Global max pooling and Global average pooling. The former converts an input of dimension (height, width, filters) to (1.1,filters) by getting the maximum value throughout the height and width dimensions for every filter. Global average pooling acts in the same way but it calculates their average value.

With this layer, we can determine the output size of our neural network to match it with the number of classes to train a classification problem regardless of the input size. It is important to note that there is always a minimum size per image with is derived from all of the filter size throughout the neural network and the padding or stride parameters. A schema of a vanilla, Fully Convolutional Neural Network is represented in Fig. 1.

*2.3. Audio description*

Two techniques were employed to describe the audios: Mel spectrogram and Mel Frequency Cepstral Coefficients (MFCC). These techniques are described below.

*2.3.1. Mel spectrogram*

The Mel scale is the result of non-linear transformations of the frequency scale. Eq. (1) represents the transformation of $f$ Hertzs into $m$ mels.

$$m = 1127,01048\log_e(1+f/700) \qquad (1)$$

The Mel scale describes the perceived spacing of frequencies. The reference point between this scale and the frequency scale is defined by equating a 1,000 Hz tone, 40 dBs above the listener's hearing threshold, with a 1,000 mels tone. Above 500 Hz, exponentially spaced frequency intervals are perceived as being linearly spaced. Consequently, four octaves on the hertz scale above 500 Hz are compressed into about two octaves on the Mel scale. The name Mel comes from the word melody to indicate that it is based on the human perception of tones.

Therefore, Mel spectrogram is a spectrogram with the Mel scale as its *y* axis.

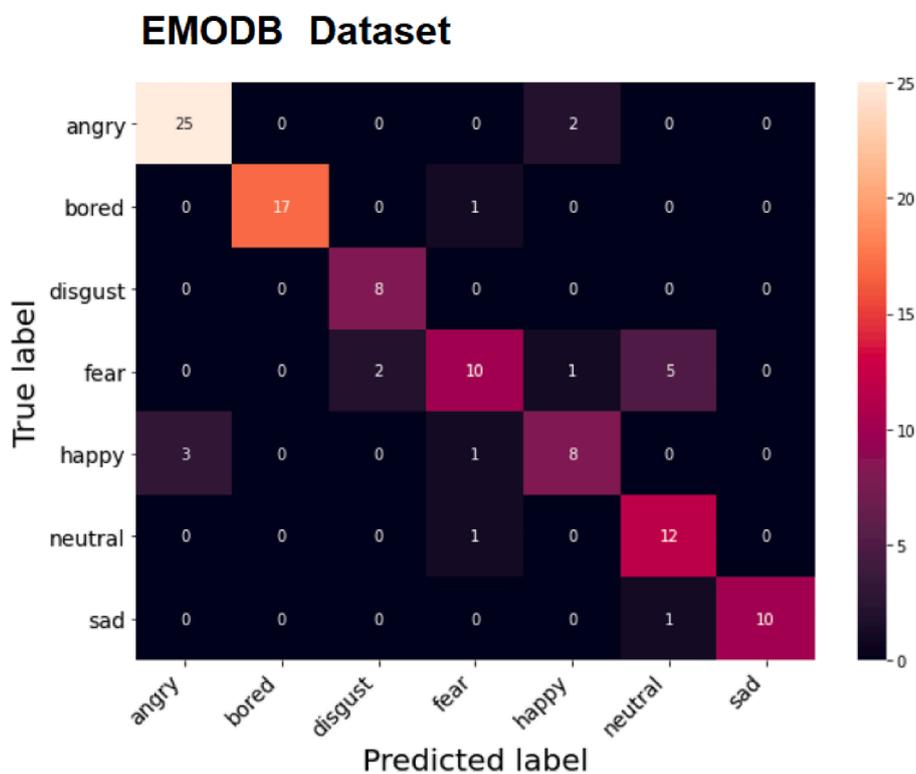

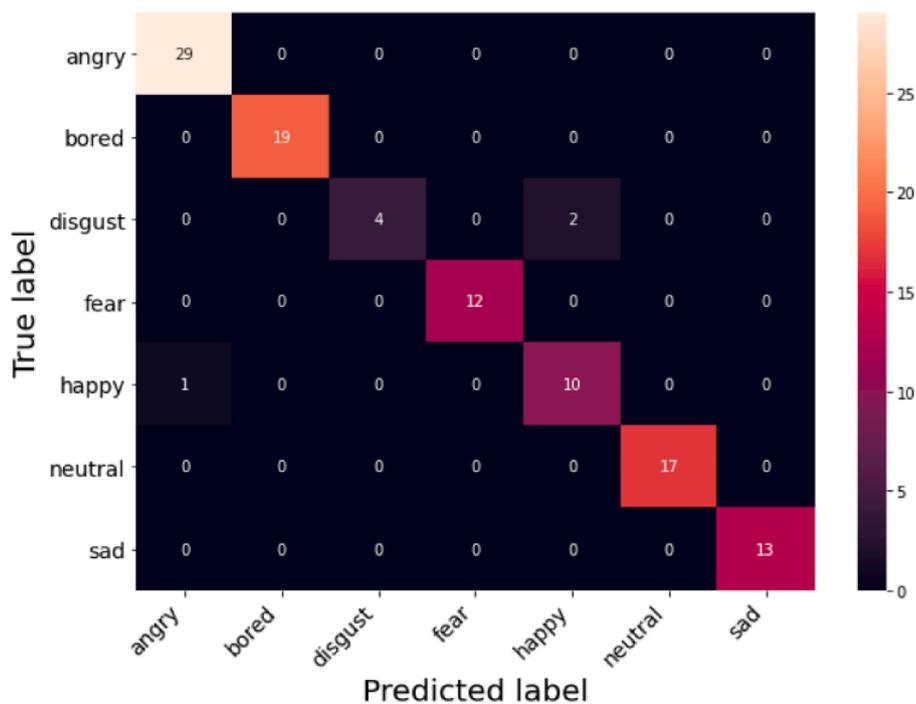

Fig. 10. EMODB confusion matrix for the best training with (a) mel spectrogram audio description and (b) MFCC audio description.

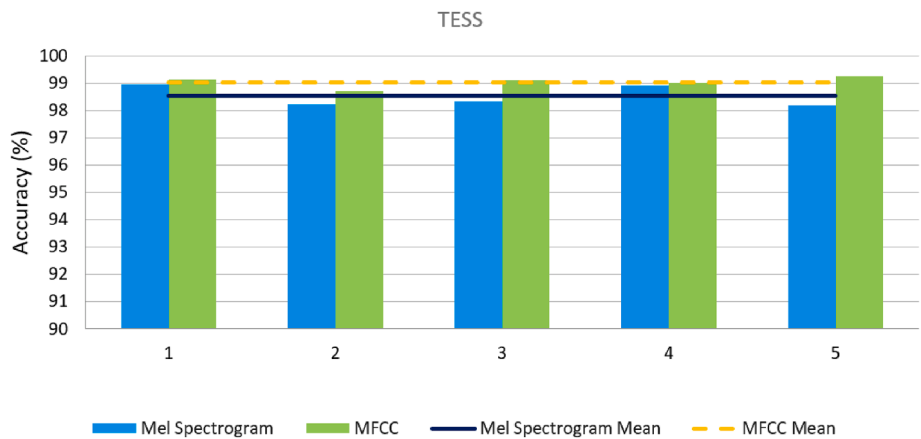

**Fig. 11.** TESS accuracy for each of 5 test subsets and their mean using Mel Spectrogram and MFCC audio descriptions.

**Table 3**
TESS Results summary.

| Audio Descriptor | Mean | Median | Std |
|---|---|---|---|
| Mel Spectrogram | 98.52 | 98.34 | 0.3691 |
| MFCC | 99.03 | 99.09 | 0.2095 |

### 2.3.2. Mel Frequency Cepstral Coefficients

Mel Frequency Cepstral Coefficients or MFCCs are coefficients for the representation of speech based on human auditory perception. These arise from the need, in the area of automatic audio recognition, to extract the characteristics of the components of an audio signal that are suitable for the identification of relevant content, as well as to obviate all of those that have little valuable information such as background noise. MFCC coefficients represent the width of the spectrum of the speaks compactly. This has made them the most widely used feature extraction technique in speech recognition [47]. The calculation of the MFCC includes the following steps: segment the audio file into small sections, apply the discrete Fourier transform (DFT) to each section and obtain the spectral power of the signal, apply the bank of filters corresponding to the Mel Scale to the spectrum obtained in the previous step and add the energies in each of them, take the logarithm of all the energies of each Mel frequency, and finally, apply the discrete cosine transform (DCT) to these logarithms and finally the first N coefficients are selected. The entire process is summarized in Fig. 2.

## 3. Experiments and results

### 3.1. Experimental setup

#### 3.1.1. Data preprocessing

The audio files obtained from RAVDESS [44], EMODB [45] and TESS [46] datasets were described in two ways: Mel spectrogram and Mel Frequency Cepstral Coefficients (MFCC). Both audio descriptors were calculated using the librosa Python library. In the case of the Mel spectrogram, the sample rate (the number of samples per second) is 22,050, the length of the Fast Fourier Transform (FFT) window is 2,048, the number of samples between successive frames is 512 and the exponent for the magnitude melspectrogram is 2 (representing the power).

In the case of MFCC, the first 100 coefficients have been selected experimentally. The sample rate is 22,050 like in the Mel spectrogram description and the normalization used is the orthonormal discrete cosine transform.

In both cases, the audio files have been processed without cuts or other transformations. This is possible due to the intrinsic FCN properties which allows variable input sizes. This also allows us to split the full length audio file into sub samples to achieve a near real time sentiment analysis.

#### 3.1.2. Network architecture

A Fully Convolutional Neural Network has been implemented from scratch.

The decision to use this type of architecture is established because the main aspect that is dealt with in this work, is twofold: evaluate a full sentiment audio and split it in smaller audio files to carry out an almost "real time" tracking of the sentiments and distinguishing the sentiments of every speaker in a conversation independently of the length of each utterance.

The configuration details are as follows:

FCN is made up of convolutional layers, three in this case. The first one contains 64 filters and a kernel size of (7.11). The second one is made up of the same number of filters and a kernel size of (11.7). Both convolutional layers use the relu activation function. As the network is an FCN, the third convolutional layer has as many filters as classes and the kernel size is equal to one. After this third convolutional layer, a Dropout layer has been appended to avoid overfitting. A Global Average Pooling layer has been added to calculate the average output of each feature map in the previous layer with the aim of reducing the data and preparing the model for the final classification layer using a Softmax activation function.

With this architecture, the output size of the neural network is determined to fit it with the number of classes to train a classification problem regardless of the input size.

The architecture scheme is shown in Fig. 3.

As we can see in the Fig. 4, the FCN architecture converts each input shape into a fixed output size, which allows us to perform the classification regardless of the duration of the audio file. The key factor of this network is the GlobalAveragePooling layer that extracts a unique value from each filter that corresponds to the average value of all filter weights. For that reason, it does not matter how long the audio file is. This is very relevant in evaluating long audio files to identify a general emotion in a long speech or to identify emotion in near real time dividing the audio file in smaller ones.

#### 3.1.3. Training setup

A Monte Carlo cross validation has been carried out in order to ensure generalization. We have split the dataset into 5 different subsets with an 80–20 distribution for training and testing. Accuracy has been used as the performance metric to evaluate the different approaches.

The intertwining of hyperparameters, and the time-consuming procedure for network training, finding an efficient network configuration for CNN is a challenge [48]. In this work, the same neural network has been trained throughout all of the experiments using Adam as an optimizer and categorical crossentropy as the loss function. The FCN was trained over 10,000 time periods with an early stop callback to avoid overfitting and a batch size of 80, obtained experimentally.

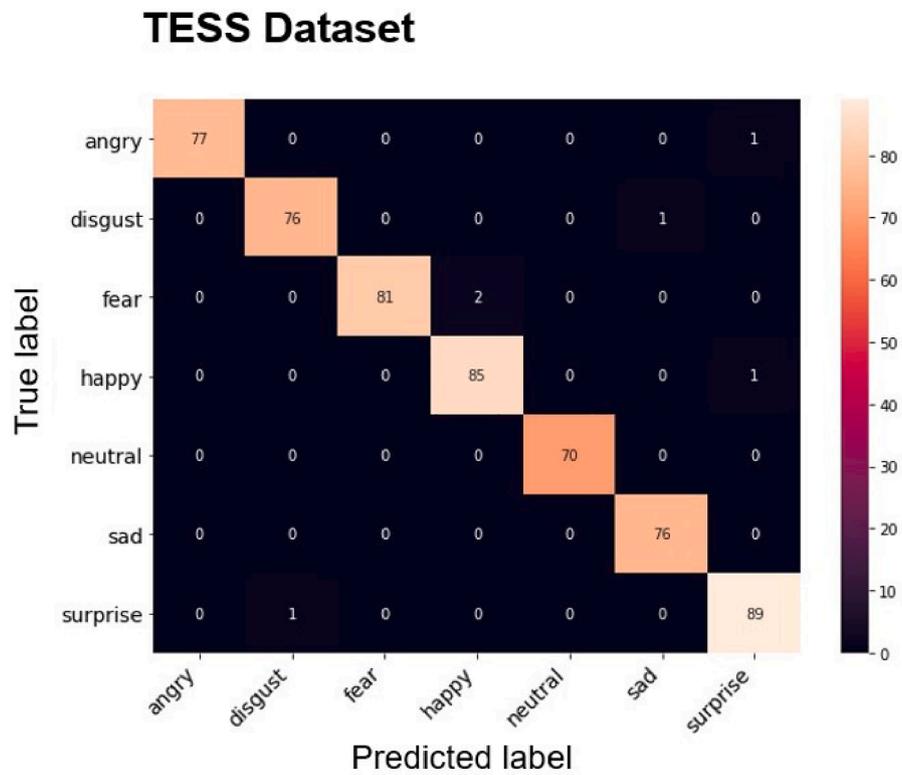

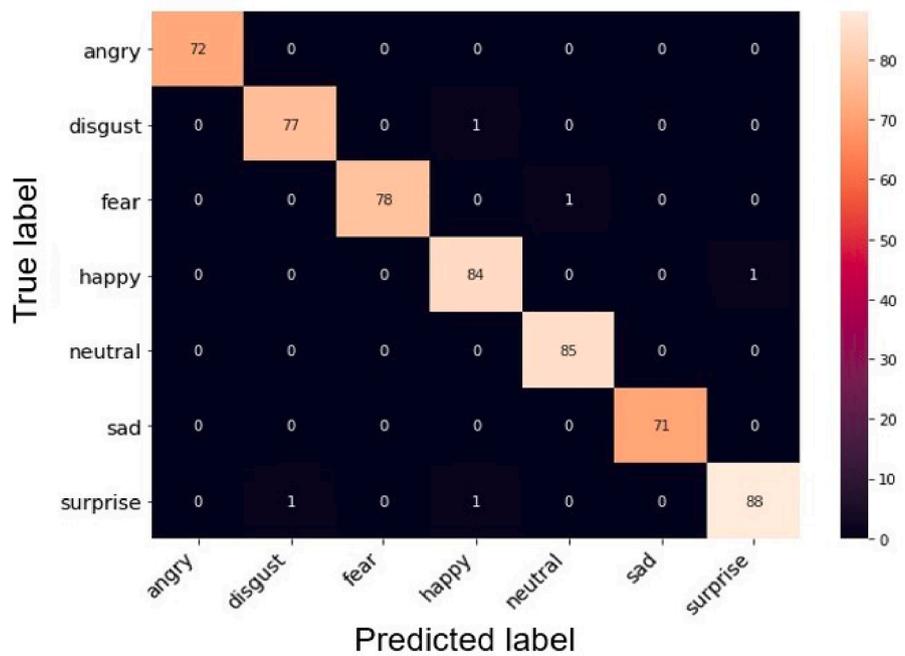

Fig. 12. TESS confusion matrix for the best training with (a) mel spectrogram audio description and (b) MFCC audio description.

**Table 4**
State of the art comparison on RAVDESS dataset. Accuracy and Improvement rate with each proposal.

| Paper | Accuracy (%) | Improvement (%) |
|---|---|---|
| Shegokar and Sircar [49] | 60.10 | 25.26 |
| Zeng et al. [50] | 65.97 | 14.11 |
| Livingstone and Russo [44] | 67.00 | 12.36 |
| Parry et al. [51] | 65.67 | 14.63 |
| Jalal et al. [52] | 68.10 | 10.54 |
| Issa et al. [1] | 71.61 | 5.12 |
| **Our proposal** | **75.28** | 0.00 |

**Table 5**
State of the art comparison on EMODB dataset. Accuracy and Improvement rate with each proposal.

| Paper | Accuracy (%) | Improvement (%) |
|---|---|---|
| Badshah et al. [53] | 52.00 | 78.29 |
| Wang et al. [54] | 73.30 | 26.48 |
| Lampropoulos and Tsihrintzis [55] | 83.93 | 10.46 |
| Huang et al. [56] | 85.20 | 8.81 |
| Wu et al. [57] | 85.80 | 8.05 |
| Issa et al. [1] | 86.10 | 7.68 |
| Mustaqeem et al. [24] | 85.57 | 8.34 |
| Anvarjon et al. [23] | 92.02 | 0.75 |
| Seo et al. [28] | 86.92 | 6.66 |
| Mustaqeem et al. [25] | 90.01 | 3.00 |
| **Our proposal** | **92.71** | 0.00 |

**Table 6**
State of the art comparison on TESS dataset. Accuracy and Improvement rate with each proposal.

| Paper | Accuracy (%) | Improvement (%) |
|---|---|---|
| Dupuis et al. [46] | 82.00 | 20.77 |
| Praseetha et al. [58] | 95.82 | 3.35 |
| Huang et al. [59] | 85.00 | 16.51 |
| Zafar Iqbal et al. [60] | 97.00 | 2.09 |
| Patel et al. [22] | 96.00 | 3.16 |
| **Our proposal** | **99.03** | 0.00 |

Furthermore, the confusion matrix has been calculated to identify the strengths and weaknesses for the different classes.

In Fig. 5 we can see the chartflow of the training step. In our case, each audio file is labeled in the filename using the first letters of the emotion. For each audio files the mel spectrogram representation or the MFCC features are extracted. As we can see, the length of the audio files can be variable thanks to the Fully Convolutional Neural Network design. The training was performed and the resulting model is stored for inference.

In Fig. 6 we can see the inference process, for both the offline or online alternative. In the offline process, the input of an audio file of any length is stored. The complete audio file is described by the MFCC or Mel spectrogram and the stored model obtained in the training step is used for the prediction of emotions. In the online process, the audio emitted by a speaker is processed every second (or any other small interval of time) and an emotion is predicted having, at the end, a sequence of emotions which allows us to obtain a better interpretation of the complete audio file if it is too long.

### 3.2. Results

#### 3.2.1. RAVDESS

Mel Spectrogram and MFCC description methods have been used to describe the data before training the proposed FCN. The results for each of the 5 iterations on the Monte Carlo cross validation can be seen in Fig. 7.

With this dataset, both description methods achieved similar results, MFCC being better in 3 out of 5 iterations. In Table 1 the mean, median and standard deviation is shown. As we can see, the MFCC outperforms the Mel Spectrogram in 1.5.

Furthermore, using the best classifier for each description, a comparison of their confusion matrices is shown in Fig. 8. As can be seen, sad and neutral are the classes with the worse performance in both cases while the calm class is one of the best using the Mel spectrogram and surprise is one of the best recognized classes using MFCC. Although the training showed a 75.27% of mean accuracy, the classifier learned the main characteristics of each class, having the most errors in similar emotions, for example, the neutral class has usually been misclassified as sad or calm.

#### 3.2.2. EMODB

In the same way, our proposal has been evaluated using a learning model based on fully connected neural networks with an EMODB dataset taking into account two different feature extraction processes: Using Mel spectrograms and MFCC features. The results achieved for each of the 5 iterations on the Monte Carlo cross validation can be shown in Fig. 9. In contrast with the RAVDESS results, MFCC showed significantly better results compared to the Mel spectrogram approach in all subsets, obtaining in the best of them an accuracy of 97.19%. The mean, median and standard deviation of all the iterations taking into account the accuracy metric can be seen in Table 2. MFCC outperforms in 17.82% the results achieved by the Mel Spectrograms considering the mean value with a 92.13% of accuracy. In this case, both techniques had a similar standard deviation throughout the different experiments.

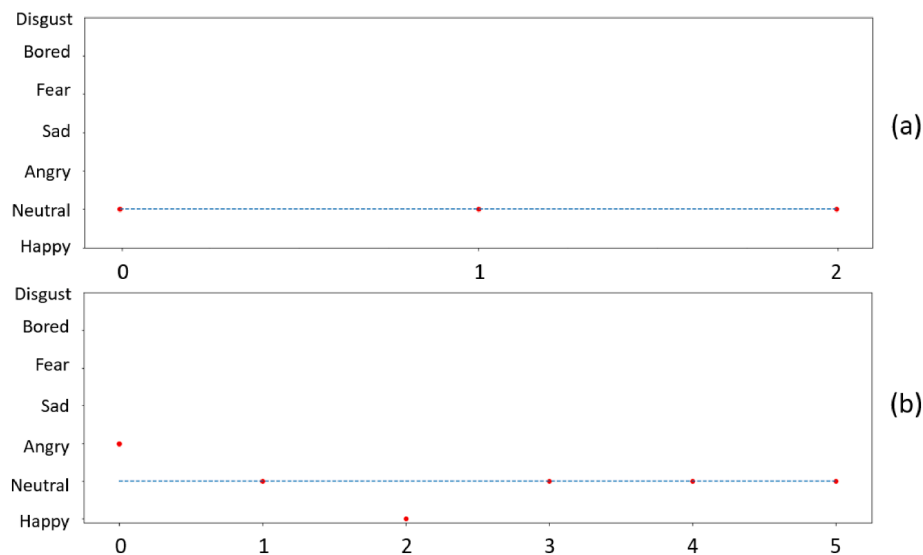

**Fig. 13.** Sentiment prediction on 3 and 6 split audio subsamples of neutral class.

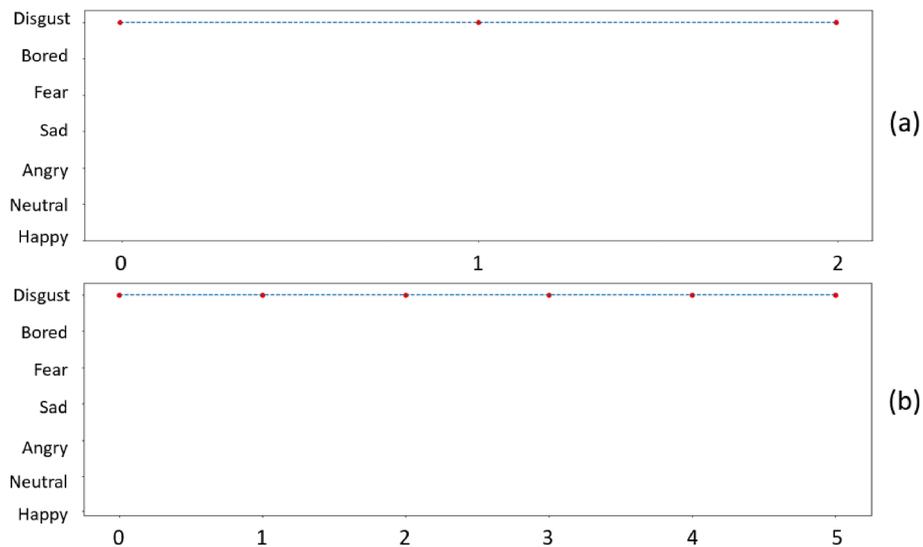

**Fig. 14.** Sentiment prediction on 3 and 6 split audio subsamples of disgust class.

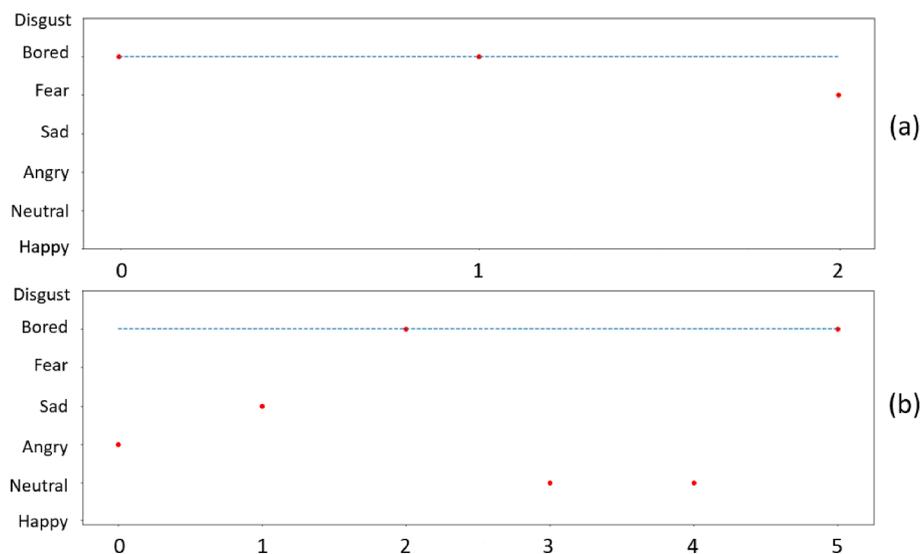

**Fig. 15.** Sentiment prediction on 3 and 6 split audio subsamples of bored class.

As with the RAVDESS dataset, a comparison between the best two classifiers has been made showing the confusion matrix of both descriptions in Fig. 10. The MFCC description only fails in two disgust and one happy audio samples. On the other hand, the Mel spectrogram only predicts the disgust samples perfectly. Here it is noticeable that with the Mel spectrogram, 3 happy samples were missclassified as angry and 2 angry samples as happy. This can be explained by the similar high tones emitted in these two kinds of emotions.

### 3.2.3. TESS

TESS dataset has been also considered for evaluating our proposal by taking two different feature extraction processes into account: Mel spectrograms and MFCC features. The results achieved for each of the 5 iterations on the Monte Carlo cross validation can be shown in Fig. 11. The mean, median and standard deviation of all the iterations taking into account the accuracy metric can be seen in Table 3. In this case, both methods achieve a high performance close to 99% of accuracy, the MFCC results being a bit higher (99.03%) than the Mel Spectrogram ones (98.52%). In this case, both techniques had a similar standard deviation throughout the different experiments. These results demonstrate the good performance of the proposal throughout all of the datasets evaluated.

As with the other datasets, a comparison between the best two classifiers has been made showing the confusion matrix of both descriptions in Fig. 12. The MFCC description only fails in five cases showing a perfect performance in sad and angry classes. Mel Spectrogram has a similar performance but with only perfect results for neutral class.

### 3.3. State of the art comparison

In order to evaluate the performance of our proposal, we have compared it against the state of the art. In Table 4, we can see that the RAVDESS results were achieved by some of the most recent works in the state of the art. Our proposal based on FCN obtained the best results with 75.28.

In Table 5, some of the most recent results of emotion recognition using the EMODB dataset can be seen. Our proposal outperforms all of the methods studied with a 92.71.

In Table 6, a comparison with the state of the art with the TESS dataset can be seen. Our proposal outperforms all of the methods studied with a 99.03% of mean accuracy after cross validation.

In addition to the results, it is also important to point out that our proposal is the only one that allows audio files of variable duration.

### 3.4. Real time evaluation

One of the main advantages of using a Fully Convolutional Neural Network is the capacity of the net to predict emotions in different length of inputs. This allows us to create a near real time sentiment analysis to check not only the main emotion but also secondary emotions that can appear throughout the audio. Two different experiments have been carried out dividing each original audio file from the EMODB dataset into three or six different splits. With this, we have been able to identify emotions in audio files of approximately one second and 0.5 s respectively. Some examples of the near real time evaluation results are shown in Figs. 13–15. Red points represent the emotion detected in each split and the blue line represents the general label of the audio.

In Fig. 13, a neutral audio file has been evaluated showing how, when it is split into 6 segments, 4 out of 6 are detected as neutral but an angry and another happy audio interval is identified.

Fig. 14 showed a disgust audio which is correctly classified as disgust in every split. In Fig. 15, we can see an example of a boredom in the audio. After splitting it into three subsamples, 2 out of 3 are marked as bored. However, when we split it into 6 segments, there is a lot of variety, which can lead us to think that this audio is very ambiguous or that it is even made up of different voices.

### 4. Conclusions

In this paper, a new neural network architecture for emotion recognition in audio has been proposed. A fully convolutional network (FCN) has been developed, firstly, to deal with emotion classification in three well-known datasets (RAVDESS, EMODB and TESS) and secondly, to enable near real time sentiment analysis to be able to analyse the evolution of a conversation, which is really interesting for numerous enterprises such as banks, call centers or even hospitals. Two different feature descriptions have been made in order to identify the best approach: Mel Spectrograms and MFCC features. The results have shown how, in all datasets, the MFCC features outperformed the Mel Spectrograms when treating them as an image. When evaluating the RAVDESS dataset, our proposal with MFCC achieved 75.28% of mean accuracy, which represents an improvement of more than 5% over state-of–the-art techniques. Furthermore, the same architecture (FCN) with the same feature description (MFCC) on the EMODB dataset has 92.71% of mean accuracy, which represents an improvement a 0.75% over recent research. In the same way, over TESS dataset our proposal achieved an accuracy of 99.03%, outperforming the state of the art in a 2.09%

Besides the exceptional results achieved, our method is able to process variable input lengths which enables it to classify near real time splits in the audio with promising results. Furthermore, the proposal has achieved the same performance in different languages as RAVDESS audios are in English while EMODB are in German, which makes it very useful for multi-language analysis.

In future works, a more profound evaluation of different audio emotions will be carried out by merging different datasets in order to be able to distinguish a wider number of sentiments.

**Funding**

This work was supported by the Spanish Government (Ministerio de Economía y Empresa - Secretaria de Estado para el Avance Digital) in 2019 with the reference TSI-100909–2019-64.

**CRediT authorship contribution statement**

**María Teresa García-Ordás:** Conceptualization, Formal analysis, Investigation, Methodology, Software, Validation, Writing - original draft. **Héctor Alaiz-Moretón:** Methodology, Writing - review & editing. **José Alberto Benítez-Andrades:** Conceptualization, Methodology, Software, Writing - review & editing. **Isaías García-Rodríguez:** Formal analysis, Investigation, Validation, Writing - review & editing. **Oscar García-Olalla:** Formal analysis, Investigation, Validation, Writing - review & editing. **Carmen Benavides:** Investigation, Validation, Writing - review & editing.

**Declaration of Competing Interest**

The authors declare that they have no known competing financial interests or personal relationships that could have appeared to influence the work reported in this paper.